# Freeze-dried microfluidic monodisperse microbubbles as a new generation of ultrasound contrast agents

Ugur Soysal,[1,*,†] Pedro N. Azevedo,[1,2,3,*] Flavien Bureau,[2] Alexandre Aubry,[2] Marcio S. Carvalho,[3] Amanda C. S. N. Pessoa,[4] Lucimara G. De la Torre,[4] Olivier Couture,[5] Arnaud Tourin,[2] Mathias Fink,[2] Patrick Tabeling[1]

[1] Microfluidique, MEMS et Nanostructures, IPGG, ESPCI Paris, Université PSL, CNRS, France;
[2] Institut Langevin, ESPCI Paris, Université PSL, CNRS, France;
[3] Department of Mechanical Engineering, PUC-Rio, Brazil;
[4] School of Chemical Engineering, University of Campinas, Brazil;
[5] LIB, Sorbonne Université, CNRS, INSERM, France;

\* These authors contributed equally to the work.
† Corresponding author: email ugur.soysal@espci.fr

**Abstract -** In the paper, we succeeded to freeze-dry monodisperse microbubbles without degrading their size and acoustic properties. We used microfluidic technology to generate highly monodisperse (coefficient of variation, CV<5%) microbubbles and optimized their formulation along with a cryoprotectant. By using a specific technique of retrieval of the bubble, we showed that freeze-drying the microbubbles does not alter their size distribution (CV≤6). To compare the fundamental resonance properties of the bubbles, we performed backscattered acoustic characterization measurements. Our experimental results revealed that the freeze-drying process conserved the acoustic properties of the bubbles. The maximum backscattering power amplitude of fresh and freeze-dried monodisperse PVA bubbles was around ten eight times higher than that of SonoVue at a similar concentration *in vitro*. By solving the question of storage and transportation of monodisperse bubbles, our work facilitates their penetration in the domain of UCAs, for performing new tasks and developing novel non-invasive measurements, such as pressure, unaccessible to the existing commercialized bubbles.

## INTRODUCTION

Since the first use of free gas bubbles as ultrasound contrast agents (UCAs) in 1968 (Gramiak and Shah 1968), significant efforts have been made to improve UCAs' stability, safety, and echogenicity. Today, UCAs are stabilized with a shell that encapsulates a heavy gas produced by sonication and agitation of a solution of surfactants; lipids, proteins, polymers. (Frinking et al. 2020). Current manufacturing techniques succeed in producing the appropriate structure but, on the other hand, provide limited size control of microbubbles, leading to polydisperse contrast agents, typically from 1 to 10 µm in diameter (Stride et al. 2020).

Polydisperse microbubbles have a clear and practical advantage: the same product covers low and high frequencies, enabling the image of the deep or shallow structure, respectively. This way, a single product can have multiple uses. However, limitations induced by size polydispersity have been underlined over the last years. In polydisperse UCAs, the resonance frequency of the microbubbles being, according to Minnaert law, inversely proportional to their size (Minnaert 1933), the resonance spectrum will be broad (Tremblay-Darveau et al. 2014) (Hoff et al. 2000). Ultrasound scanners used in clinics deliver acoustic pulses with a narrow frequency bandwidth, leading to the excitation of only a small portion of the bubbles near their resonance in a polydisperse population (Segers et al. 2016). Thus, polydispersity is not optimal, from the signal-to-noise ratio perspective (Stride et al. 2020); it also hampers to fully benefit from the non-linear characteristics, which enables minimizing shadowing effects in deep tissue imaging (Segers et al. 2018b).

Moreover, the use of polydisperse bubbles prevents performing quantitative measurements that could be relevant for clinical applications, such as measuring local pressure (Fairbank and Scully 1977) (Ishihara et al. 1988) (Tremblay-Darveau et al. 2014) (Segers et al. 2018a). Local pressure affects, according to Minnaert law, the resonance frequency. Should we have a monodisperse population of bubbles, there could be a pathway leading from resonance measurements to local pressure. However, polydisperse bubbles produce a broad spectrum, from which it remains difficult to extract quantitative information on the pressure. For example, the required clinical sensitivity for portal vein hypertension diagnosis is (±2 kPa), and the measurements are expected to be reproducible with a monodisperse population, namely with a coefficient of variation (CV) of 6% (Tremblay-Darveau et al. 2014). Attempts using subharmonics were also proposed (Shi et al. 1999) (Esposito et al. 2020), but from a purely empirical basis that may weaken the reliability of the measurement.

In molecular diagnosis, the bubbles were functionalized to develop an affinity with specific antigens (Leong-Poi Howard et al. 2003). In the bloodstream, the bubbles preferentially settle in specific areas, revealing the presence of the targeted molecules. The number of bubbles reaching their targets is low and thus generates weak echoes (Frinking et al. 2012). The high sensitivity of resonance frequency shift on the discrimination of freely circulation and bound bubbles is a necessary condition not met by polydisperse bubbles (Overvelde et al. 2011).

Microbubbles were also used to transport drugs, genes, or therapeutic gases (Hernot and Klibanov 2008) (Sutton et al. 2014). When bubbles reach specific areas, they can locally release the charge by bursting. Nevertheless, polydisperse bubbles require high ultrasonic intensities to achieve efficient bursting, which can cause clinical problems. Thus the use of monodisperse bubbles provides controlled and efficient release (Roovers et al. 2019). Therapeutic applications such as sonothrombolysis (Dixon et al. 2019) and blood-brain opening (Choi et al. 2010) can also be optimized owing to the precise control of monodisperse bubbles.

Recently, phospholipid-shelled microfluidic monodisperse microbubbles were characterized with 2-3 orders of magnitude higher acoustic sensitivity than that of polydisperse microbubbles *in vitro* (Segers et al. 2018b). Likewise, the acoustic sensitivity of monodisperse bubbles was 10 times in rats and 15 times in pigs higher than that of polydisperse counterparts, such as SonoVue, Definity, and Optison (Helbert et al. 2020).

From these studies, one may conclude that monodisperse bubbles in clinical applications are highly advantageous. In some cases, monodisperse bubbles open new avenues in diagnostics, delivery, and therapy. Highly monodisperse bubbles are currently produced with microfluidics. However, to transport and store them, they need to be freeze-dried without degradation of their properties, which, today, represents a challenge. In this case, uncontrolled dynamics generated in the freeze-drying process induce rearrangement, coalescence, and ripening phenomena that degrade monodispersity. In fact, freeze-drying without degradation is the major bottleneck to unlock, allowing monodisperse microfluidic bubbles in clinics.

In the past, alternatives to freeze-drying were proposed. It consisted of, first, producing monodisperse droplets with ink-jet printing (Böhmer et al. 2006) or microfluidics (Song et al. 2018), using oil-water or air-oil-water emulsions (Lee et al. 2011). These droplets were either freeze-dried or evaporated to obtain gas-filled microbubbles coated with polymer shells. All these techniques require, at some point, a toxic solvent (e.g. dichloromethane, cyclododecane, dodecane) to be removed. This procedure is a delicate task, which is never guaranteed to be fully completed, raising regulatory issues. Besides, the current polydisperse bubbles were size-filtered in different manners such as mechanical filtration (Streeter et al. 2010), centrifuge (Feshitan et al. 2009), or acoustic sorting (Segers et al. 2018b) and obtained relatively narrow size distributions, between CV of 10% to 20%. To date, there are no clinically approved either size-filtered bubbles or a size-filtering device available. Although the filtration method may be sufficient to optimize the bursting phenomenon in drug delivery or therapeutic applications, highly monodisperse bubbles, CV<6, are required for non-invasive pressure measurements. Thus, microfluidic production is a necessity. Therefore, we are left with the only pathway to freeze-dry shell encapsulated monodisperse microfluidic bubbles without using any solvent.

In this article, we succeeded in unlocking the bottleneck. We produced freeze-dried microfluidic-generated monodisperse microbubbles in a wide beneficial range that can benefit in different fields, including domains pertinent for UCA applications (from 4 to 50 µm in diameter) nanometric PVA shell-protected, without degrading monodispersity. We thus propose here a new generation of contrast agents, in the form of a stable lyophylisate, that can be stored for months and transported at any place, be resuspended, and used directly in clinical applications.

## MATERIALS AND METHODS

*Microbubble formulation*

Among the shell materials, polymer-shelled microbubbles such as poly(vinyl alcohol) (PVA) as UCAs holds an excellent potential due to their high echogenicity (Grishenkov et al. 2011) and stability (Oddo et al. 2017) in comparison to lipid-based UCAs; the material is FDA approved, and series of preclinical tests showed no toxicity (Cerroni et al. 2018). Perfluorohexane ($C_6F_{14}$) or sulfur hexafluoride ($SF_6$) was used as the gas phase of the system and thereby the bubbles filled with $C_6F_{14}$ or $SF_6$. Such gases are highly hydrophobic, less soluble, and heavier than air in aqueous solutions and were selected in the second generation microbubbles (Podell et al. 1999), (Schneider 1999). As the permeable shell leads to a rapid escape of soluble gases, such as air, the use of heavy and less soluble gaseous enhances the lifetime of the bubbles. The liquid phase is an aqueous solution of polyvinyl alcohol (PVA) (Mw 13,000-23,000, 87-89% hydrolyzed, Sigma-Aldrich) at a concentration of 8 wt. % in 10 mM Phosphate-Buffered Saline (PBS) solution in Milli-Q water. PVA was stirred in the aqueous solution at 80°C overnight for being used as the continuous phase. The surfactant allows for the formation of a shell around the bubbles; thus, PVA acts as a surfactant and shell material. To facilitate the freeze-drying of the bubbles, a cryoprotectant is required to avoid the destruction of the bubbles during the harsh and freeze-drying process, which is associated with freezing, dehydration, and solid-liquid interfacial stresses (Abdelwahed et al. 2006), (Date et al. 2010), (Mensink et al. 2017). The external fluid includes both the cryoprotectant and surfactant. In our novel easy-to-prepare formulation, PVA acts both as a surfactant and a cryoprotectant. Although PVA is not currently used as a cryoprotectant for freeze-drying of microbubbles, it received attention as an emerging cryoprotectant for droplets (Song et al. 2018) and pharmaceutical products (Mitchell et al. 2019).

*Description of the device and production of the' fresh' microbubbles*

The device shown in Fig. 1 was realized with soft lithography technology (Tabeling 2005). In short, the mold was fabricated using double-layer conventional lithography. PDMS (PolydiMethyl Siloxane) was poured into the mold and baked at 70°C for 30 min. Then, the inlet and outlets were punched through the PDMS, and the microfluidic channel was sealed with a microscopic glass slide by applying $O_2$ plasma with a plasma cleaner (The CUTE, Femto Science Inc., Republic of Korea). Finally, plastic tubings were connected to the inlet and outlet of the device.

The device of Fig. 1 contains two inlets for liquid injection, two outlets for the recovery of excessive material (i.e., anti-clogging system), one gas inlet, and one bubble outlet for collecting the generated bubbles. The bubble outlet consists of a lateral tube aligned along the axis of the collection channel to avoid coalescence phenomena (e.g., lateral collection). It also includes an entry (e.g., pusher) that allows to increase the flow rate of the continuous phase and tune the bubble concentration. This step favors the high throughput and reliability of the process. The bubbles with 5 µm in diameter were produced with the flow-focusing method. The width and height of the flow-focusing geometry are 5 µm. This structure is accompanied by a step-emulsification geometry that enables enlarging the structure with an upward step-like channel (height of 20 µm) to reduce the hydrodynamic resistance. An anti-clogging system (so-called river channels) was included to avoid large aggregates or contaminants clogging the channels to ensure a prolonged stable generation (Malloggi et al. 2010). Note that the large bubbles, such as 44 µm, were generated with a typical flow-focusing device without adding the river channels to the design as the nozzle diameter larger than 20 µm is usually less prone to clogging.

The liquid and gas flows were pressure-driven, $p_{liquid}$, and $p_{gas}$, using an MFCS-100 pressure controller system (Fluigent) at pressures from 1.2 to 3.5 bars. In turn, also taking into account the aforementioned geometrical features of the device, the microfluidic system enabled us to generate 5 µm in diameter microbubbles (CV <5%) with up to 10 kHz generation frequency, relatively high throughput (~$6\times10^5$ bubbles/min) (see Fig. 1). In situ microbubble generation was recorded using a high-speed camera (Photron SA3, Japan)-connected optical microscope (Zeiss Observer A1, USA) at 20x and 40x magnifications. The recordings were analyzed to obtain the size distribution and the generation rate of the bubbles in MATLAB.

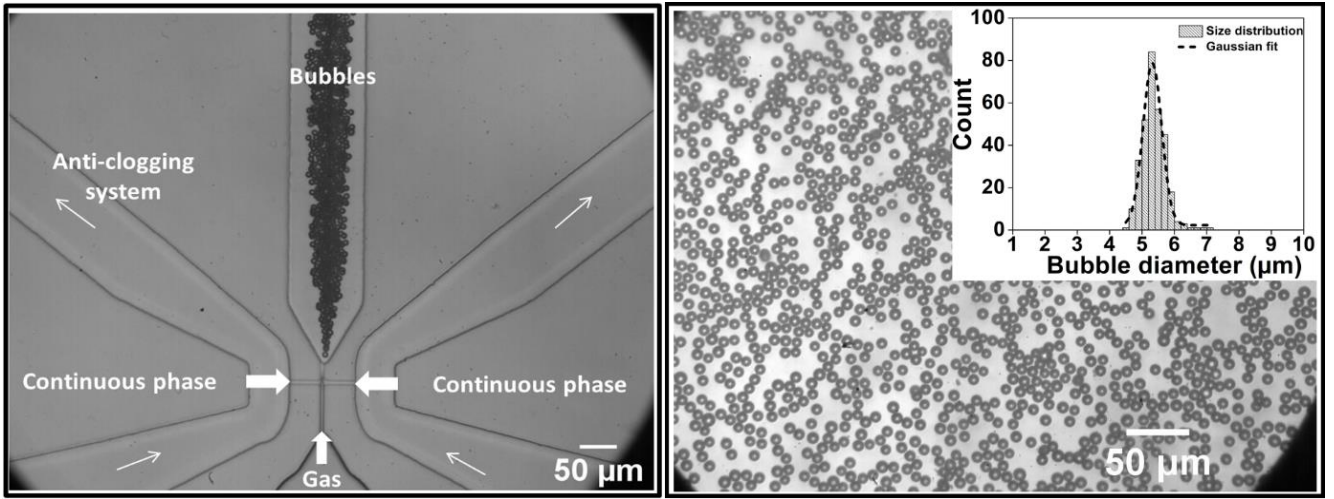

Fig.1. Microfluidic generation of 5 µm in diameter PVA 'fresh' microbubbles (i.e., directly coming out of the microfluidic device and thereby *not* resuspended from a lyophilisate). (Left) Monodisperse bubbles were generated in the central channel located between river channels and after the gas channel. Bubbles exited the device without any coalescence. (Right) Freshly generated bubbles were collected in a monolayer manner on a glass slide. The inset shows the size distribution of the collected bubbles showing average sizes, standard deviation, and CV are 5.4 µm, 0.3 µm, and 5.5%, respectively.

*Method of acoustic characterization of microbubbles*

The acoustic characterization setup, instrumentation, example of typically emitted signals, received echoes, and resonance spectra for the backscattered acoustic measurements of nearly stationary microbubbles are shown in Fig. 2. We used a small PDMS-based structure as a measurement chamber (35mm×10mm×1mm). The thickness of the walls of the PDMS container is 10 mm, more than ten times larger than the initial wavelength. This ensures propagation of the initial wave deep in the system and produces echoes from the container walls. The structure was replicated with PDMS from a mold.

A focused ultrasound transducer (with a central frequency of 2.25 MHz, a relative bandwidth of ~50%, a quality factor of ~2, and a 38-mm focal length) was used as an emitter and a receiver of echoes (see Fig. 2A). The Minnaert resonance frequency for an uncoated 5.4 µm in diameter bubble is around 1.2 MHz and calculated as follows (Minnaert 1933):

$$f_o \times d_o = 6.6 (MHz \times \mu m) \qquad (1)$$

Where $f_o$ is the resonance frequency and $d_o$ is the diameter. A shell-coated bubble is expected to resonate to ultrasound at a frequency higher than the Minnaert resonance frequency. For thin polymer-shelled microbubbles, where the shell thickness was 5% of the bubble radius, the linearized Church model (Church 1995) was developed by (Hoff et al. 2000). In this model, the shell increases the resonance frequency by increasing the stiffness through the shear modulus while the ratio between shell thickness and bubble radius is assumed to be constant during the bubble oscillation (see Eq. 2).

$$f_o = \frac{1}{2\pi d_r} \sqrt{\frac{1}{\rho_L}\left(3\kappa\rho_o + 12 G_s \frac{d_{Se}}{d_r}\right)} \qquad (2)$$

Where $d_r$ is bubble radius, $\rho_L$ is the density of the liquid, $\kappa$ is polytropic constant, $\rho_o$ is the hydrostatic pressure in the surrounding liquid, $G_s$ is the shear modulus, and $d_{Se}$ is shell thickness. Based on this model, (Hoff et al. 2000) estimated the shell shear modulus of thin polymer-shelled UCAs between 10.6 and 12.9 MPa. Later on, (Grishenkov et al. 2009) estimated the shear modulus of PVA bubbles ranging between 3.5 MPa and 5.7 MPa based on a similar approach that (Hoff et al. 2000) developed. We may assume that the shear modulus of our PVA bubbles will be in the same order of magnitude. Based on the Church-Hoff model, we analytically estimated that the expected resonance frequency of 5.4 µm in diameter monodisperse PVA bubbles (shell thickness of ~70 nm, see Results section) with the shear modulus from 3 to 12 MPa results in 2.1 MHz to 3.7

MHz that lies within the bandwidth of the transducer which covers the frequencies from ~1.1 MHz to ~3.4 MHz. Thus, the selected transducer is compatible and may directly target the resonance frequency of the monodisperse PVA bubbles.

We used a function generator (Tektronix AFG 2021) to generate a Sinusoidal 1-cycle burst signal with a 1-ms pulse repetition period (see Fig. 2B). The peak negative acoustic pressure, as measured by a hydrophone (see Fig. 2C), was fixed to 134 kPa. A PicoScope (5242D) was used for data acquisition, and its dedicated software was used to monitor and record the detected signal. It acquired 1 waveform/45ms and one waveform (i.e., echo) consists of 3125 data points. Typically, each measurement presented here took 45s, yielding 1000 waveforms. Each echo was taken from the focal point, as the region of interest (i.e., ROI typically consists of 350 points and is around 60µs from the initial wave due to the delay of the propagation in the solid PDMS structure). Its Fourier transform is then calculated (via an FFT algorithm) and averaged with a MATLAB program.

When there was no bubble but water in the measurement chamber, the obtained spectra consisted of reflections from the measurement chamber (see Fig. 2B and Fig. 2C). Therefore, a metallic plate was placed in the chamber, rather than placing it in a big water tank, to obtain the transducer response without underestimating the present experimental boundaries. Thereby, the spectra, consists of the transducer response along with the reflections from the chamber, were obtained with the metallic plate. Then, it was deconvoluted from the spectra obtained with the bubbles. The deconvolution process was performed by dividing the bubble spectra (see Fig. 2D, left) by the metallic plate spectra (data not shown). The result of the division was fitted with a Gaussian to analyze the resonance frequency along with the standard deviation (see Fig. 2D, right).

The resonance measurements were compared as a function of a number of microbubbles semi-quantitatively determined from $10^8$ to $10^6$ with dilution ratios of 1:1 to 1:100 for SonoVue in up to 500 µl liquid. Although the concentration of SonoVue was roughly estimated, it still gives an insight for the comparison with monodisperse bubbles. The number of freshly generated 5 µm in diameter bubbles (same bubbles as shown in Fig. 1) was estimated based on the collection time, namely 20min to 30s, to obtain approximately $10^7$ to $10^5$ bubbles. Note that obtaining a higher number of bubbles than $10^7$ with our current device caused long waiting times, making the generated bubbles prone to coalescence in a vial. PVA bubbles were diluted with PFC saturated PBS solution to enhance the stability and gently injected into the measurement container. Although the injected liquid volume was smaller than that of the container, it remains trapped and intact between the walls due to the capillarity and therefore did not spread through the channel; thus, air bubbles were avoided (see Fig. 2A).

Fig. 2B shows the obtained echoes from the measurement chamber, filled with Milli-Q water or PBS (see Fig. 2A), without the presence of the bubbles. The echoes produced from the front wall of the PDMS container, from the front and back (merged echoes) walls of the measurement chamber, and from the back wall of the PDMS container are denoted as X, Y, and Z respectively. Our various experiments revealed that echoes produced by the front and back wall (i.e., merged echoes) of the measurement chamber define the detection limit of our measurement setup. As shown in Fig. 2C, the merged echo (shown as Y in Fig. 2B) is one order of magnitude smaller than that of the bubbles, which guarantees a good signal-to-noise ratio. The separation of these merged echoes is possible and splits the echo from the bubbles and the reflections (see Fig. 2B, echo Y). However, it requires increasing the channel height that either makes the measurements prone to multiple scattering (which is neglected since absorption dominates over scattering (Tourin et al. 2000) (Hoff 2001)) or leads to freely floating bubbles in the chamber. Lowering the concentration of the bubbles, e.g., down to $10^3$ to $10^2$ bubbles, leads the measurement to fall below the detection limit in our setup while freely floating bubbles easily escape from the focal region hampering repetitive measurements from the same bubble population over time. This optimized geometry enabled us to robustly demonstrate and compare the backscattered acoustic response of freshly generated and freeze-dried monodisperse microbubbles. Typical results for a population of monodispersed 'fresh' bubbles, i.e., retrieved from a microfluidic generator, such as in Fig. 1, are shown in Fig. 2C and Fig. 2D.

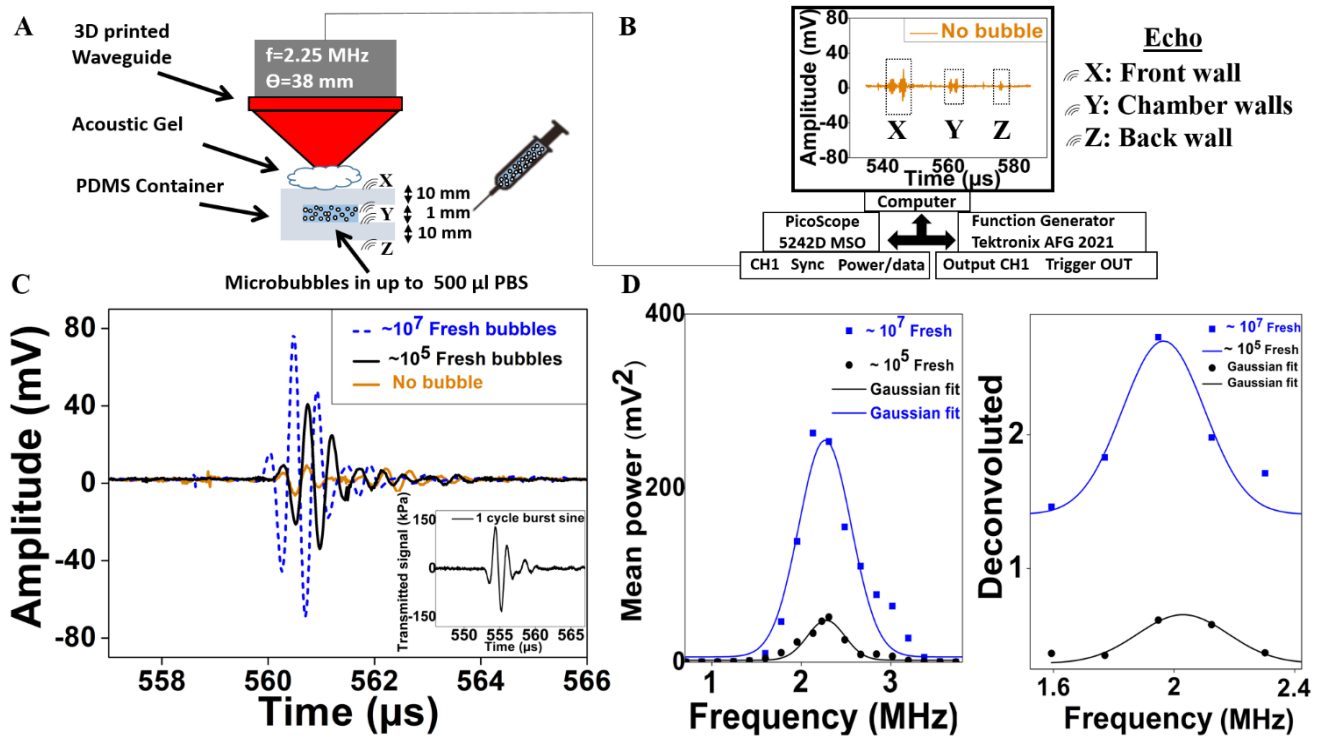

Fig.2. Acoustic characterization setup and typical acoustic response of fresh bubbles. (A) The acoustic measurement test bench illustrates a focused transducer with an acoustic gel-filled 3D printed waveguide and a PDMS container where the bubbles were injected in up to 500µl water. (B) The description of the data acquisition, monitoring, and recording of the detected signal with a PicoScope (5242D) is shown. It provided an acquisition rate of 1 waveform/45ms, which consists of 3125 data points. An arbitrary wave function generator (Tektronix AFG 2021) generates a Sinusoidal 1-cycle burst signal with a 1 ms pulse repetition frequency. (C) The backscattered echoes show an acoustic comparison of ~$10^7$ and ~$10^5$ fresh PVA monodisperse microbubbles with a reference measurement (where there are no bubbles). The slight misalignment between echoes originated from the alignment of the transducer to the channel through the acoustic gel in different experiments. The inset shows an emitted signal with peak negative acoustic pressure of 134 kPa. (D) (Left) The typical resonance spectrum (average of 1000 spectra) of ~$10^5$ and ~$10^7$ fresh bubbles with a ~%10 standard deviation from the Gaussian center is shown. (Right) The deconvoluted spectra (described in materials and methods section) of ~$10^5$ and ~$10^7$ fresh bubbles show in the order of ~%10 standard deviation from the Gaussian center, for both.

*Monolayer bubble collection and freeze-drying protocol*

The process is shown in Fig. 3. The generated bubbles were transported along with the lateral outlet tubing and exited the device. Then, the bubbles formed milky suspension drops at the tip of the tubing. Each 10 µl drop was spread on an $O_2$ plasma-treated hydrophilic glass slide (75 mm x 25 mm). Thus, the bubbles were collected in a monolayer manner, resulting in individual localized bubble suspension spots on a glass slide (see Fig. 3).

We typically obtain one hundred monolayer spots on the slide, each of which consists of $10^4$ bubbles with 5 µm in diameter. Each spot results from the deposition of a single droplet (<10 µl). Following the monolayer collection, the bubbles were pre-freeze for a minimum of 2h at –25°C, far below the glass transition temperature (Tg) of the used cryoprotectant (~80°C). Finally, frozen bubbles were transferred to a freeze-dryer (FreeZone Labconco, USA) with a shelf temperature of -50°C for 4h. Each freeze-dried spot then was hydrated with a gas saturated PBS solution on the glass slide in a PFS ambient. Should monodisperse microbubbles be collected in large drops leading to the formation of multilayers, as we did for comparison, bubbles would interact before being frozen, the lyophilisate would be multilayer, and the resuspended micro-bubble population polydisperse.

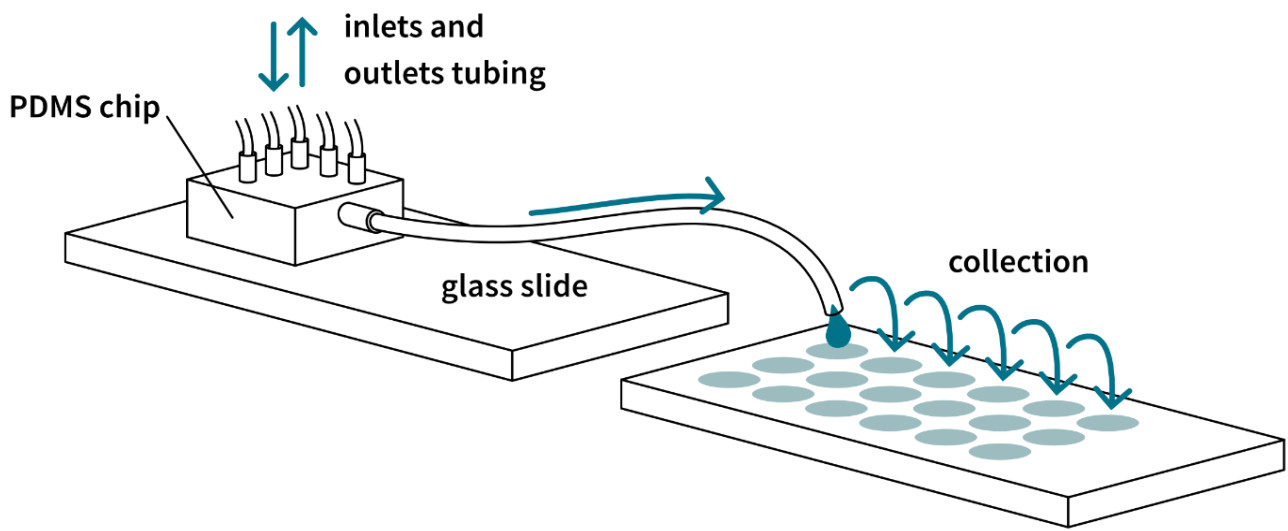

Fig.3. An illustration of the microfluidic production of bubbles in a drop and spot-based drop collection on a glass slide.

## 3. RESULTS

*Geometrical characterization of the freeze-dried bubbles*

The lyophilisates that we obtained, in the process described in Material and Methods, were uniform in size, as shown in Fig. 4A and Fig. 4B. In Fig 4A, the bubble diameter is 44 µm. They form a closed-packed structure. This is because, with the concentrations at hand, the system stands above the jamming transition (Furuta et al. 2016). Increasing the volume fraction of the bubbles leads to close-packed states, as shown in Fig. 4A. Whereas increasing the volume fraction of the liquid by decreasing the size of the same number of bubbles unjam the system, as shown in Fig. 4B. The size distribution, measured with a Matlab program, shows a monodispersity of 10%, thus substantially larger than the 6% dispersion of the initial fresh bubble population, but still low. The reason originates from the fact that freeze-dried bubbles interact and undergo deformation due to the close packing. This behavior also indicates that, by forcing bubbles to be collected as a single monolayer and freeze-dried in such a configuration, even though in-plane interactions exist, they slightly degrade the size uniformity but do not destroy the monodispersity. However, the origin of in-plane interactions is not fully understood. Unlike 44 µm bubbles, in Fig. 4B, bubbles are smaller (5 µm), concentration is lower, and the systems stand below the jamming conditions. Consequently, the lyophilisate is a distribution of well separated, isolated, 5 µm bubbles.

The environmental scanning electron microscope (ESEM) image shown in Fig. 4C was taken with a 20° tilt, without metallic coating, to avoid modifications on the shells, but resulted in one side go white due to the charging effect. The corresponding ESEM images show non-spherical bubbles (as might be expected because there is no surface tension), the shells look uniform, suggesting acceptable homogeneity of the lyophilization conditions across the population. The thickness of the shell was estimated as ~70 nm from a broken shell, shown as an inset in Fig. 4C, a non-tilted and non-coated image, which is consistent with (Poehlmann et al. 2013), who reported PVA shell thicknesses ranging between 120 nm and 230 nm.

Fig. 4D and Fig. 4E show the same populations after rehydration and resuspension in DI water. The detachment from the glass slide and, thus, resuspension of the bubbles took only a few seconds after they were hydrated. Then, the bubbles started to freely float in water. The size distribution was measured when they were hydrated at the same plane, but it also remains unchanged during the floating. Fig. 4D and Fig. 4E show that the size distributions of 44µm and 5µm bubbles are 6% and 5.5%, respectively, comparable to the fresh bubbles, showing that monodispersity of the size is accurately conserved during the process. This is the main result of the paper. Note that while 44µm bubbles shrunk by 20% in size after freeze-drying, 5µm bubbles shrunk only a few percent.

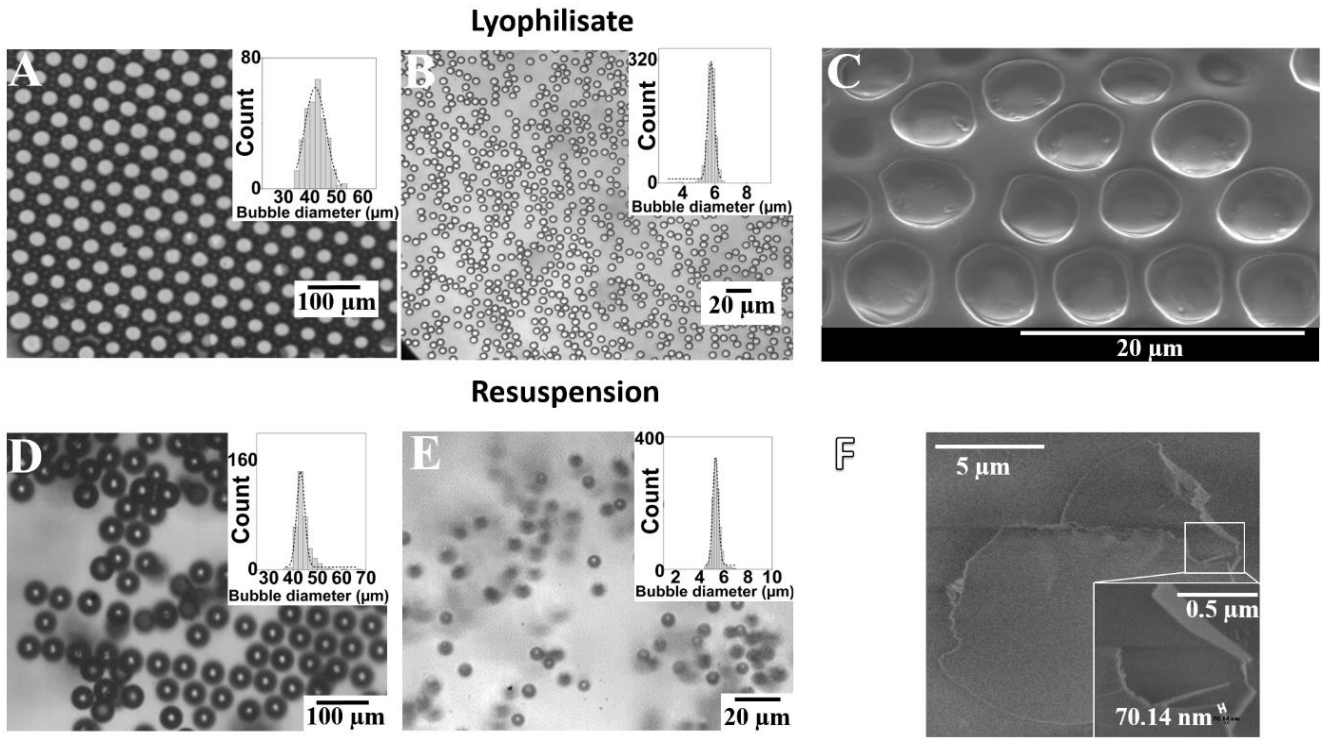

Fig.4. The images and size distribution (bars represent the size distribution and dots represent Gaussian fits) of lyophilisates and resuspended bubbles. (A-B) The optical images of the central part of lyophilisate spots consist of freeze-dried bubbles deposited onto glass slides. The insets show the size distribution of the lyophilisates 42 µm (standard deviation of 4 and CV of 9%) and 5.8 (standard deviation of 0.3 and CV of 5%) µm in average diameters of the lyophilisates. (C) ESEM of the 5 µm lyophilisate. (D-E) Resuspended and floating bubbles and their size distributions (insets) show the average diameters of the bubbles as 44 µm (standard deviation of 3 and CV of 6%) and 5.4 µm (standard deviation of 0.3 and CV of 5.5 %). (F) SEM image shows a broken bubble, and the magnified image measures the thickness of the shell as ~70 nm.

*Acoustic characterization of resuspended freeze-dried bubbles*

The main objective for performing acoustic characterization of freshly generated and freeze-dried monodisperse PVA bubbles was to investigate possible effects of the freeze-drying process on the physico-chemical properties of the bubbles, which would ultimately alter the fundamental oscillating properties of the bubbles, such as resonance frequency and its amplitude. Therefore, we performed *in vitro* backscattered acoustic measurements, both for freeze-dried and freshly generated bubbles, under the same experimental conditions. These experiments were performed in the setup described in Fig. 2A at the similar number of bubbles (~$10^6$), with the same size (5µm as shown in Fig. 1 and Fig. 4E), same gas core (PFC), and same surrounding liquid (PBS) to uncover the consequence of freeze-drying process only on the oscillating behavior of the bubbles by acoustical excitation. Fig. 5A shows, in dashed red and dashed black lines, that the acoustic responses of the freeze-dried and fresh bubbles are indistinguishable. Thus, we may suggest that the freeze-drying process does not alter the viscoelastic properties of the bubbles. In any case, from a UCA perspective, we may conclude that freeze-dried and fresh bubbles behave, within experimental error, in the same manner. The conclusion is confirmed in the spectral space. Fig. 5B shows the resonance spectra, in solid red and black, (both were average of one thousand measurements) of before and after freeze-dried bubbles.

We observed identical responses, within ±2% in the resonance frequency and ±10% in the maximum amplitude. Likewise, Fig. 5C shows, in solid red and black, the same data, but deconvoluted according to the process shown in Material and Methods. Here again, the response of freeze-dried and fresh bubbles, within experimental error, are undistinguishable. The number of data points that constitute the spectra was limited due to the resolution limit of our data acquisition. Still, their Gaussian form is clear. The resonance responses of both fresh (black curve) and freeze-dried PVA bubbles (red curve) were characterized with around 10% standard deviation, ensuring a narrow bandwidth of monodisperse bubbles. This finding acoustically confirms the aforementioned optical image-based results that the monodispersity was kept intact until the resuspension of the bubbles. The resonance frequency we infer from Fig 5C, from the gaussian fit, is 2.01 ± 0.04 MHz. From the measured resonance frequency of the fresh and freeze-dried bubbles, 2.01±0.04 MHz, we may analytically estimate the shear modulus of the bubbles, around 2 MPa, based on the Church-Hoff model (Hoff et al. 2000). This is in good agreement with (Grishenkov et al. 2009) who analytically and experimentally reported the shear modulus of the PVA bubbles was in the same order of magnitude. On the other hand, Minnaert law (Minnaert 1933) or more elaborate models (Doinikov and Bouakaz 2011) suggest that the size distribution of the bubbles, such as 6% in CV, translates into a 6% standard deviation in the bandwidth of the resonance response. However, it is known that the oscillating bubbles are subjected to several damping mechanisms along with the influence of the shell, the occurrence of reflections from the experimental setup that we observed, all leading to a broadening of the resonance peak (Hoff et al. 2000).

It is interesting to compare our PVA bubbles to SonoVue commercial contrast agents. In Fig. 5A, the solid blue line represents the acoustic response of SonoVue at a similar concentration, and it was around five times less than the response of the fresh and freeze-dried PVA bubbles. We also found that the maximum backscatter power amplitude response of monodisperse freeze-dried PVA bubbles was about ten times greater than that of SonoVue (data not shown). The larger overall echogenicity of monodisperse bubble populations has been established by (Segers et al. 2018b) for a number of commercial contrast agents (Helbert et al. 2020). The explanation relies on the fact that in polydisperse bubbles, with the standard excitations employed, only a fraction of them respond, while in the monodisperse population, all respond when the applied ultrasound frequency matches with the resonance frequency of the bubbles. We confirmed this feature with our freeze-dried microfluidic bubbles.

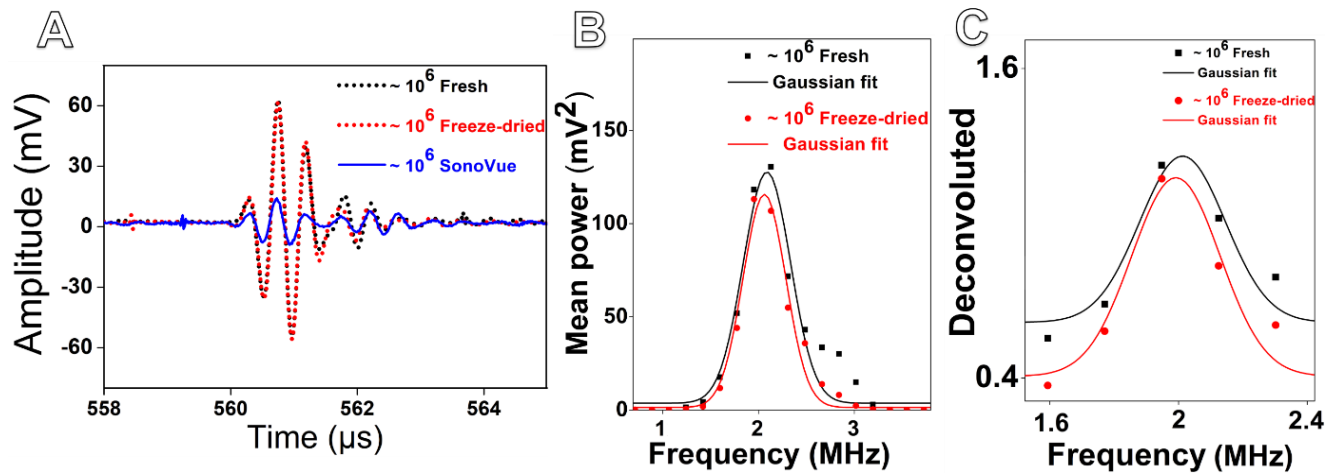

Fig.5. Typically obtained backscattered echo and the resonance spectra. (A) The echoes of fresh and freeze-dried PVA monodisperse bubbles with SonoVue at a similar concentration are shown. (B) The resonance spectra of the fresh and freeze-dried same size of 5µm in diameter PVA bubbles (both are average over one thousand spectra) show indistinguishable backscattered acoustic responses (the difference of ±2% in the resonance frequency and ±10% maximum amplitude) at a similar concentration and under same experimental conditions. (C) The deconvoluted resonance spectra of both fresh and freeze-dried PVA bubbles showing around ~%10 standard deviation from the Gaussian center with a difference of ±1% in the resonance frequency and ±10% maximum amplitude at a similar concentration.

## DISCUSSION

Many progress on the stable production of monodisperse microbubbles as UCAs and their numerous applications were successfully performed until today. Based on the evidence in the scientific literature, while the monodispersity improves the performance of microbubbles in ultrasound image quality, it can make it possible for bubbles to be used in early diagnosis, treatment, and sensing applications. The polydispersity of current clinical UCAs hampers the use of bubbles in such emerging applications. But, thus far, no technology has enabled the storage and transportation of monodisperse microbubbles. The alternative solutions to freeze-drying are not realistic compared to the clinical UCAs, which are storable and transportable, and easy to use by non-experts. Beyond commercial and regulation aspects, this may have slowed down the advancement of monodisperse bubbles through the clinics, despite its superior qualities, concerning several pertinent aspects, in comparison to commercial polydisperse bubbles.

The successful freeze-drying strategy is composed of two steps. The first is the optimization of microbubble composition and buffer solution formulation, including selecting a shell molecule, a surfactant, and a cryoprotectant. While the shell and surfactant improve the stability of the bubbles, cryoprotectant protects the bubbles during freezing and freeze-drying cycles. In our case, PVA acts as shell material, surfactant, and cryoprotectant. Second, monolayer and localized freeze-drying of microbubbles guarantee a monodisperse population through the resuspension of the bubbles. The latter leads to a reduced interaction among the bubbles. Therefore, the failure of step one leads to destroying the bubbles after freeze-drying, while step two is leading polydisperse freeze-dried bubbles.

Noteworthy to investigate a number of properties of our lyophilisates along with our resuspended bubbles that can be relevant for applications. In this spirit, we tracked the resonance frequency of resuspended PVA bubbles over a few minutes, using the acoustic setup of Fig. 2A. This time represents billions of oscillations. We did not observe any significant change in the resonance frequency. This behavior suggests acceptable shell physico-chemical stability since it is known that the shell constitution substantially contributes to the resonance frequency determination (de Jong et al. 2002). In addition, it turned out that the lyophilisates were transported around the city to perform acoustic measurements with our collaborators, and we did not observe any degradation in the acoustic performances. Last but not least, the lifetime of the lyophilisate is at least one year. This was obtained by using a lyophilisate stored at ambient conditions at room temperature for one year. In this study, we found that neither morphologies, diameter distributions, nor resonance frequencies significantly evolved during storage.

It must be mentioned that the microfluidic technology is capable of producing monodisperse microbubbles tunable in size, e.g. down to 1 µm in diameter (Gnyawali et al. 2017). The small bubbles are to be addressed to *in vivo* applications on small animals where high frequencies (e.g. 10-15 MHz) are typically preferred due to the limited penetration depth. To facilitate the use of large sizes of microbubbles on the small animals, radial modulation imaging can be used to detect large microbubbles, e.g. 5 µm in diameter, at high frequency by using dual ultrasonic excitations such as 1 MHz and 15 MHz (Muleki-Seya et al. 2020).

Contrarily to what is sometimes thought, using microfluidic devices for producing monodisperse bubbles does not jeopardize the possibility of reaching throughputs compatible with research and clinical needs on a world scale. Rough estimates of world productions of contrast agent microbubbles are on the order of $10^{17}$ per year. A microfluidic device working continuously at a frequency of 10 kHz, by avoiding perturbation in the production with the anti-clogging system (see Fig. 1), produces $10^{11}$ bubbles per year. By parallelizing one thousand times, this would lead to $10^{14}$ per year. Besides, obtaining 1 million microbubbles per second from a single microfluidic device would only lead to the use of a few devices (van Elburg et al. 2021). If we add that when monodisperse bubbles are used, one hundred times fewer bubbles are needed, owing to their high echogenicity (see Fig. 5A) and (Segers et al. 2018b), (Sirsi et al. 2010), (Helbert et al. 2020) the production we can realize with microfluidic devices is compatible with the needs of the domain, on a world scale. Indeed, before freeze-drying, we need to spread the bubbles on surfaces, which may be space-consuming. However, calculations show that, in an industrial process where bubbles are produced, then brought to the freeze-drier, the area needed to spread the bubbles would be extremely limited.

## CONCLUSIONS

For the first time, we freeze-dried monodisperse microbubbles without degrading their monodispersity. This strategy unlocks the bottleneck of transportation and storage of monodisperse microbubbles in the form of a lyophilisate, i.e., a powder. We solved this problem by developing two different aspects. The first is the optimization of bubble formulation along with a cryoprotectant. The second is maintaining the bubbles, during collection, pre-freezing, and freeze-drying steps, in the form of monolayer spots deposited on a glass surface. Although we developed this method on a polymer-shelled microbubble, in principle, our method can be applied for obtaining freeze-dried monodisperse lipid or protein-shelled microbubbles.

In our experiments, fresh and freeze-dried PVA bubbles, having the same size and gas core, were acoustically characterized under the same experimental conditions. In such conditions, we observed, within experimental error, the same backscattered acoustic response. These measurements complete the optical observations of sizes and morphologies across the freeze-drying process and lead to conclude that freeze-drying, in the way we proceed, conserves the bubble properties.

Moreover, the lyophilisates were stored over a year-long time at room temperature and transported to other laboratories around the city to perform acoustic measurements. We did not observe any geometrical nor acoustical degradation due to the storage and transportation of the lyophilisates. Moreover, we compared the backscattered acoustic response of SonoVue and monodisperse PVA bubbles and found that the maximum acoustic backscattered response of monodisperse bubbles was around ten times higher than that of SonoVue bubbles at a similar concentration. Our work might facilitate the penetration of monodisperse bubbles in the domain of UCAs, for performing new tasks and developing novel non-invasive measurements.

*Acknowledgments - We gratefully thank the Institut Pierre Gilles de Gennes (IPGG), CARNOT IPGG, ESPCI, CNRS, PSL-Valorisation, and Coordenação de Aperfeiçoamento de Pessoal de Nível Superior – Brasil (CAPES- Institutional Internationalization Program – Print) for their financial support. We are grateful to Elian Martin, Gilles Renault, Etienne Coz, Pierre Garneret, Maria Russo, Igor Braga de Paula for discussions and help.*